# On the Fundamentality of Meaning[1]

Brian D. Josephson

Department of Physics, University of Cambridge, Cambridge CB3 0HE, UK

**Abstract**

The mainstream view of meaning is that it is emergent, not fundamental, but some have disputed this, asserting that there is a more fundamental level of reality than that addressed by current physical theories, and that matter and meaning are in some way entangled. In this regard there are intriguing parallels between the quantum and biological domains, suggesting that there may be a more fundamental level underlying both. I argue that the organisation of this fundamental level is already to a considerable extent understood by biosemioticians, who have fruitfully integrated Peirce's sign theory into biology; things will happen there resembling what happens with familiar life, but the agencies involved will differ in ways reflecting their fundamentality, in other words they will be less complex, but still have structures complex enough for what they have to do. According to one approach involving a collaboration with which I have been involved, a part of what they have to do, along with the need to survive and reproduce, is to stop situations becoming too chaotic, a concept that accords with familiar 'edge of chaos' ideas.

Such an extension of sign theory (semiophysics?) needs to be explored by physicists, possible tools being computational models, existing insights into complexity, and dynamical systems theory. Such a theory will not be mathematical in the same way that conventional physics theories are mathematical: rather than being foundational, mathematics will be 'something that life does', something that sufficiently evolved life does because in the appropriate context so doing is of value to life.

---

[1] FQXi essay competition entry. An addendum includes material added subsequent to the submission of the essay.

**Introduction**

Recently I gave a talk at a conference on fundamental physics, entitled 'Incorporating Meaning into Fundamental Physics'. The definition of fundamental physics I had in mind in proposing that title is the one that most physicists probably have, the main issue being the *universality and wide-ranging nature* of the theories involved. In this connection, the theory known as the Standard Model fits a wide range of phenomena, but cannot take into account gravity, while on the other hand a different theory, that of General Relativity, gives a good account of gravitation, but that is all it is capable of. Under these circumstances neither theory can be considered fundamental, so particle physicists seek a more universal, and hence more fundamental, 'theory of everything'. My thesis is that such theories can equally not be considered fundamental, since they fail to take proper account of the phenomenon of *meaning*. Meaning fails to show up in the world of physics simply because the kind of situations that physicists prefer to investigate are ones where meaning has no significant influence on the outcome, a situation analogous to that of weak interactions, which can very often be ignored but which play an essential role in the case of phenomena such as beta-decay; in both cases we have a situation where an important phenomenon does not feature in everyday physics because it is generally irrelevant there. The conclusion to be drawn is that the kind of 'theory of everything' currently hoped for by particle theorists would in reality be merely 'theories of everything that physicists are interested in'.

The reason why meaning is normally considered non-fundamental is that it is regarded as being *emergent from a deeper level*, but others have disagreed. David Bohm for example (Bohm 1987) has asserted that 'meaning is capable of an indefinite extension to ever greater levels of subtlety', implying presumably that there is more to meaning than is commonly understood by scientists. Even if this is the case, would that have any significance for the world of physics? In this regard, it might be argued that thoughts are influenced by the subtleties of meaning referred to by Bohm, and at the same time have observable effects that current physical theories do not take into account, implying that they are inexact[1]. It is not enough however, for the purposes of physics, just to say 'something interesting seems to be happening here'. In the case of weak interactions, genuine progress was made by showing how existing theories could be extended to include these interactions. A similar kind of extension would be desirable for meaning also, so as to lead to full integration of this entity into fundamental physics.

I will argue in the following for the likely possibility of such an extension, based however on *biology* rather than physics. The aspect of biology involved is that of *biosemiotics*, the application to biology of semiotics, the name given to the nineteenth-century theory of signs due to the philosopher Charles Sanders Peirce. This modern approach to the understanding of biological complexity examines the way the effective functioning of specific systems depends critically on the ability to make appropriate responses to relevant categories of information, in other words taking into account the *significance* of such information in regard to current activity. In this way, one comes to understand many aspects of the structure and processes of biological systems.

---

[1] In this connection a case can be made as in Josephson and Carpenter (1996a), based upon an objective analysis of regularities discernible in the corpus of musical compositions, that musical aesthetics involve subtleties not currently accommodated within science.

**Relationships between the biological and quantum domains**

Biology is normally considered not relevant to fundamental physics, but its possible relevance is suggested by correspondences between the quantum world and that of biology such as those shown in the table of fig.1, taken from a paper entitled [Beyond Quantum Theory: A Realist Psychobiological Interpretation of Physical Reality](#) (Conrad et al. 1988):

| LANGUAGE OF QUANTUM PHYSICS | | LANGUAGE OF BIOLOGY |
|---|---|---|
| quantum subsystem, describable by a state vector | ↔ | signal or form |
| particle type | ↔ | type of signal or form |
| state vector representing a specific possibility | ↔ | signal representing a specific possibility |
| collapse of state vector | ↔ | decision process |
| measuring instrument determining state of subsystem | ↔ | structures which determine and regulate signals or forms |

It was suggested in the paper that such correspondences might stem from the existence of a common underlying mechanism, which would radically transform our view of the quantum realm. Similar parallels have been noted by others, as for example in the well known comment of James Jeans, who wrote in *The Mysterious Universe*: "the universe begins to look more like a great thought than like a great machine. Mind no longer appears as an accidental intruder into the realm of matter; we are beginning to suspect that we ought rather to hail it as a creator and governor of the realm of matter...". Again, John Archibald Wheeler claimed in his [Law without Law](#) that 'a principle, that of observer-participancy, might suffice to build everything'. Similarly, [in an interview](#) Wheeler said: "We are participators in bringing into being not only the near and here but the far away and long ago", while again former particle theorist Karen Barad, author of the book *Meeting the Universe Halfway: Quantum Physics and the Entanglement of Matter and Meaning* (Barad 2007) asserted, again on the basis of similar parallels, that 'life forces run through everything', [also stating](#), more dramatically, that 'Matter feels, converses, suffers, desires, yearns and remembers'. As part of her analysis, Barad invokes the relationships between *apparatus* and ideas in Bohr's account of quantum measurement. Apparatus relating specific aspects of reality corresponds closely to what biological systems do[1].

All these assertions tend in the direction of suggesting that, at some level that science does not currently understand, nature is in some sense biological. While such ideas as those cited may arouse scepticism, how much do we actually know from science in regard to what is really happening at the quantum level? Mainstream physics has as it were washed its hands of the whole business by claiming that we cannot discuss *individual* quantum events but only averages, a point of view sometimes expressed in the dictum 'shut up and calculate' (once subverted by David Mermin, who favoured instead the complementary advice, 'shut up and contemplate').

---
[1] The measuring apparatus discussed by Bohr is closely related to the *[semiotic scaffolding](#)* of biosemiotic theory.

Calculation involves mathematics, so the cited dictum implicitly presumes that nature is governed by mathematics. The biological alternative proposes here implies instead the reverse, that 'mathematics is something that nature does'[1]. But how can this alternative be properly justified? I will argue in the following that a credible theory can be created by appropriately adapting the conclusions of biosemiotics. The idea is basically quite simple: ordinary life works the way it does on the basis of certain quite subtle principles, and so at the fundamental level the same principles will apply, but will involve different actors. These biosemiotic principles will now be discussed.

**Biosemiosis**

The main concepts we shall need in our analysis are the ones involved in connection with Jesper Hoffmeyer's concept of *semiotic scaffolding* (Hoffmeyer 2008). Scaffolding is defined as something that supports specific activities, but first an important technical point needs to be clarified. Hoffmeyer relates his scaffolding concept to that of the *autonomous agents* featuring in the field of robotics. In robotics, an agent is implemented by a piece of code in a computer, and so can validly be thought of as a *thing*. Barad refers instead to *agencies* that have a similar causative role, but emphasises that they are not things but '*doings*' (or performances, or actions). The logic underlying the idea that one should talk in terms of 'doings' rather than things can be clarified by reference to the example of an earthquake. Notwithstanding the fact that grammatically 'earthquake' is a noun, the actual reference of the term is the fact that 'the earth is quaking', an *activity* with associated consequences. An earthquake is a *conceptual unit*, sufficiently common in its occurrence that languages have a word for it, generally classifying it for grammatical purposes as a noun.

This point is important because one may be tempted to view Hoffmeyer's scaffolding as a specific piece of equipment, a well-defined object, whereas in reality it is a characteristic 'doing' present in an organism. The reason why one can usefully talk about scaffolding is that, as is with the case of earthquakes, we have *models* relating to the term, which models are used in our analyses with the consequence of augmenting our understanding of the phenomena of interest to us. Biosemiotics is important for biology as its theories provide us with a window into the extreme complexity of biological systems, the relationships that it uncovers playing a role similar to those of equations in theoretical physics[1].

---

[1] The point here is that some biological systems (i.e. ourselves) have the ability to create mathematics through mechanisms that are explicable in semiotic terms. If, as argued in this presentation and also by Barad, similar mechanisms to those that allow mathematics to emerge in the human domain apply also at a fundamental level, then mathematics can equally emerge as a consequence of evolutionary processes at a more fundamental level. Accordingly, universes such as our own would then be understood as the outcome of related activity, informed by that mathematics.

We turn now to details. Hoffmeyer notes in the first place:

> *"Semiotic scaffolding operates by assuring performance through semiotic interaction with cue elements that are characteristically present in dynamic situations such as the catching of prey, invading host organisms, or mating."*

Note here the relevance of 'cue elements' (in other words signs), interaction with which is a necessity to assure *successful* performance. Specific performances are necessitated in turn by their ability to satisfy corresponding *needs* (and here we have a crucial element specific to life, clearly delimiting the complex kind of order found in living systems from the alternative kinds of order studied by the physicist). Hoffmeyer's paper goes into considerable detail regarding how the scaffolding concepts work out, as for example in the following quotes related to the question of how *new functional genes* emerge during the course of evolution. Hoffmeyer writes in this connection (such abstractions in semiotic theory being derived from careful consideration of the observed behaviour of actual systems):

> *"The decisive cause for the birth of a new functional gene would be a lucky conjunction of two events: (1) an already existing non-functional gene might acquire a new "meaning" through integration into a functional (transcribed) part of the genome, and (2) this gene-product would hit an unfilled gap in the "semiotic needs" of the cell or the embryo. In this way, a new gene may become a scaffolding mechanism supporting a new kind of interaction by imbuing some kind of semiotic advantage upon its bearer."*

and again:

> *"The emergence of new scaffolding devices (unknowingly) functions like stepping stones in a river, leading evolutionary processes forward one step at a time and — on average — farther away from the bank at each step."*

In the case of language for example, new scaffolding devices would emerge on account of their ability to fill a 'gap' in communicative capacities, this change in capacity being the product of some modification of an existing system, assisted by attention to appropriate cues related to language determined by existing scaffolding.

Further, in regard to the growth of complexity in general[2] :

---

[1] Theories constrain possibilities, as do equations, but the latter constrain in a very precise way. In the context of technology, high precision may sometimes be necessary to achieve particular aims, necessitating the use of special methodologies. Biological systems can survive without such high precision, but a degree of constraint is necessary nevertheless. While precision has its value in the biological context, high levels of precision may not be necessary for survival.

[2] I have for a long time been interested in the question of how human cognitive development occurs, and the interested reader is referred to Josephson and Hauser (1976), Josephson and Blair (1982), Josephson and Baas (1996b), and Josephson (2004). I have also been interested in the possibility of computer modelling of cognitive development, as in the computer simulation of Osborne (1995). Unfortunately, indicative of the sociological issues that sometimes confront innovative research, issues involving the department arose,    

> *"Anticipation through the skilled interpretation of indicators of temporal relations in the context of a particular survival project (or life strategy) guides organismic behavior towards local ends. This network of semiotic controls establishes an enormously complex semiotic scaffolding for living systems. "*

Rather subtle concepts are involved here, and it is helpful to compare the details of such analyses with the way large computing projects evolve since here, similarly, we have a situation where needs become apparent, and are dealt with by modifying or integrating existing mechanisms, informed interpretation playing a key role in this process. In the biological case, 'informed interpretation' emerges naturally in the course of evolution, since natural selection favours it, and thus we can not unreasonably speak of 'natural design'[1]. The key points, again, are that the principles, far from being purely theoretical, are ones closely tied to observation, and also they offer the possibility of analysis in terms of specific models.

**Adapting semiosis to fundamental levels: the Circular Theory of Yardley**

Our main strategy, as noted, involves presuming that the main concepts of biosemiotics remain applicable at a fundamental level, but applying to radically different kinds of agency. Yardley's unconventional [Circular Theory](#) fits this requirement in a number of respects. It hypotheses a specific kind of agency, the circle, acting in ways that parallel those of familiar organisms, in the first place having mechanisms serving the essential requirements, in the biological context, of survival and reproduction. The details of how they work are complicated, and only selected fragments can be detailed in this essay. A key concept is that of *oppositional dynamics*, essentially a generalisation of the familiar process of DNA replication. This involves a 'circle' consisting of two mutually supporting elements X and Y. If such mutual support exists then a situation is possible whereby X can generate Y, which then in turn generates X, and so on repeatedly, thus providing a mechanism for reproduction. But oppositional dynamics, emerging in the context of mutual reinforcement, features in many other ways, the defining characteristic being that of 'two entities acting as one'; in other words it is about coordination. And this coordination has itself a cause (given the name 'pi' on account of its relationship to the circle) which it is said 'produces stability and reliability for reality which, in and of itself, is, markedly, unstable and unreliable'. This we can understand in the following way: if an entity is capable of achieving this goal, through acts of systematically separating and joining together other systems in appropriate ways developed over time, then it is more likely to survive itself[2]. The link with biosemiotics consists in the way that pi acts as *scaffolding*, responding to cues relevant to its specific project, that of producing stability and reliability[3].

---

leading to this line of research having to be abandoned ([Josephson 2012](#)).

[1] Note that here an additional learning mechanism is necessary, since a process that was effective in the past may no longer be effective when circumstances differ. Whether a new piece of scaffolding is added on, rendering the scaffolding still more complex but at the same time effective, depends on how effective the older process is in the new situation.

[2] The competence involved in the interpretation of human language similarly involves learning through experience which constructs go together, and which do not.

[3] This suggests a connection with the 'edge of chaos', a concept featuring in discussions of complex organised systems. If a system is too chaotic it will not function reliably, but if it is not chaotic    ... cont. in footnotes on p.7

Yardley admits herself that her expositions are very confusing to readers, perhaps an inevitable consequence of the fact that she is trying to describe a situation that she herself can visualise, but which is very difficult to describe adequately in words, especially if the appropriate terminology is not available (in which connection I have benefited personally from my familiarity with the ideas and terminology of biosemiotics).

Science does however possess tools that should prove adequate to taking these ideas further, for example computer modelling (which served to disclose the existence of the previously unsuspected phenomena of chaos and the edge of chaos), dynamical systems theory, and studies involving complexity[1]. Yardley's many detailed verbal accounts can provide a good foundation for such investigations, which may then very well enhance our understanding of the natural world, interpreted as the eventual outcome of highly complex semiotic mechanisms, rather than being fully interpretable through some mathematical 'theory of everything'. Looking to the future, I suspect that a number of current dogmas, as for example the assertion that the emergence of mankind is completely accounted for by current theories of evolution, will be discredited in consequence[2]. Historians will marvel at the way insistence by the mainstream that at a fundamental level particles are the only things that matter, banishing to the fringe those scientists who thought otherwise, will be seen to have drastically interfered with the progress of science.

---

enough it will not adapt. 'Pi' would know how to balance the two requirements. In this connection, a particularly mysterious aspect of Yardley's theory is her invocation of the idea of a 'mandatory relationship between line and circle', which could be a reference to how pi achieves its goals, cycling periodically between a less chaotic state and a more chaotic one, perhaps symbolised by circle and line respectively. This relates to the familiar fact that when acquiring a skill one alternates between conditions dominated by stability, and conditions of a more innovative and unstable character, corresponding in Hoffmeyer's metaphor to the act of alternating between staying put on a particular stone, and searching for another nearby stone to step on to and then stepping on to it.

[1] An intriguing phenomenon, of possible relevance in this connection, is that of the remarkable relationships found to exist experimentally between temporal patterns (i.e. sound) and spatial ones (patterns excited by sound on the surface of water). For details see Reid 2016.

[2] In regard to *this* particular point, Yardley writes (punctuation added for clarity): "There is a symbolic man, in mind, which is the idea of man, which had to be present somewhere hidden (imaginary, an idea) before man could appear". This assertion recalls analogous facts, such as the fact that, for example, the *idea of a computer* had to be present in someone's mind before computers could come into existence. This of course can only happen because human minds have the capacity over time to transform ideas into reality. In Yardley's picture, the organising aspect at fundamental levels could evolve so as to be able to transform ideas similarly, a picture not inconsistent with Wheeler's idea of observer-participancy, and the proposals of Barad. They are, however, inconsistent with the prevailing dogma, which asserts that we live in a meaningless universe.

**Addendum, included subsequent to submission of essay**

Further reflection, in part related to [comments on the essay](#) on the FQXi website, has led to the further development of the above ideas and also in some instances their reformulation. A comment to the effect that the arguments would be better founded not on the little known semiotics but on physics led to the realisation that Peirce's *thirdness*, or equivalently triadic relationships, while it forms the basis of semiosis, has wider relevance and is indeed found in situations such as that of Jupiter's satellites, it being the case in that situation that a linear relationship exists between the three orbital phases of Io, Europa and Ganymede, given by

$$\Phi \equiv \lambda_{Io} - 3\lambda_{Eu} + 2\lambda_{Ga} = 180º \qquad (1)$$

This *physical* process involving locking together of three items presents itself as a possible mechanism underlying *semiotic* processes, also involving three items, sign, object, and the interpretant mechanism linking the two. We are concerned in that case with three systems, the third being one that under certain circumstances has the function of connecting the other two (for example, when driving).

The situation involving the satellites of Jupiter corresponds to what may be called an *established* locked-in situation. Situations can be envisaged where (1) is not satisfied even approximately, but the fact that the established situation is a stable one implies that, just as with situations involving stable equilibrium, should a situation arise such that $\Phi = 180º$ is approximately satisfied then locking in will happen, even if significant disturbances are present, such as that due to the presence of the fourth major satellite of Jupiter, Callisto. But a sufficiently large transient perturbation could disrupt a stable situation of the kind specified by (1).

The situation discussed involves in essence local order. Of more interest in this connection is what might be called global order, which as in the case of crystals involves the propagation of local order. Just as with crystals contact of local units leads to global correlation of the orientations of units, contact between individual language users leads to global correlations between the individual usages of a shared language by its users. The details of how this global order comes about lie beyond the scope of the present investigation, but the following analysis suggests a general mechanism. Ordering of the orientation of unit cells comes about as a result of adjacent cells sharing atoms and hence the relationships between atomic locations that determine the orientations of unit cells. In the same way, language users share certain features of their environment and relationships between them, and this state of affairs is responsible for the way features of a language can propagate.

In connection with the above, the hypothesised state of order associated with triadic relationships, and the crystalline state, differ radically in regard to the applicability of mathematics to their description. In the case of the crystal, the mathematics associated with crystal lattices provides a simple account of the global order, but no such straightforward equivalent exists for global order associated with local relationships such as that of eqn. (1). Nevertheless, just as happened in the case of superconductivity, where Bardeen, Cooper and Schrieffer derived a mathematical model capable of characterising the global order corresponding to the local order involving Cooper pairs, the possibility cannot be excluded that something similar can be found in the present context.

Two further points must be added to complete this conceptual analysis. The first is the fact that sharing is not the only factor determining the details of the outcome; stability issues are also important. Crystal structure will not propagate if the units themselves are unstable, and similarly relationships such as those of language, and triadic relationships in general, will not propagate if individual users do not derive stability from them. The presence of a dynamic background can lead to a situation particularly favouring systems with the highest levels of stability. In the case of language, we know that over time more and more complex organisation develops, and this we anticipate would also be possible at a fundamental level should triadic relationships prevail in this situation.

The second point is that, as noted by Yardley, a mechanism of an *oppositional* character can as effective in propagating relationships as is the process discussed above involving contact between *identical* systems. The mechanism here involves the ability of two distinct entities X and Y to form a stable relationship, as happens with the case of the two strands of DNA involved in the double helix. In such a situation, an initial structure X can organise the formation of the complementary structure Y, following which X and Y separate and Y organises the formation of its complementary structure X, thus replicating X. In the case of language two such complementary structures are the one involved in the production of speech, and the one involved with the interpretation of speech, language users needing to discover how to create structures complementary to those of the other users in their environment. Yardley applies similar arguments in connection with such matters as the way mind functions complement action structures when an action is being performed.

As discussed in the main essay, tools exist whereby detailed models incorporating the principles discussed can be investigated, and it is anticipated that over the course of time fundamental physics will develop in such a way that reality will no longer be viewed as the straightforward outcome of the application of some universal formula, but that subtle organisational issues have an equally important role to play.

**Acknowledgements**

I have learnt much from the works of Karen Barad and Jesper Hoffmeyer, and am grateful to the colleague who brought Barad's work to my attention.  I am also grateful also to Dr. Alex Hankey who has been developing related ideas (e.g. [Hankey 2015](#)), and to Ilexa Yardley, for informative discussions relating to their own approaches.